\documentclass[11pt,onecolumn]{article}
\usepackage{latexsym}
\usepackage{amsmath}
\usepackage{amssymb}
\usepackage{amsthm}
\usepackage{epsfig}
\usepackage{natbib}

\usepackage{url}
\usepackage{color}
\newcommand{\newtext}[1]{{#1}}
\newcommand{\comment}[1]{}

\usepackage{fullpage}

\begin{document}
\title{Stochastic Combinatorial Optimization under Probabilistic Constraints}
\author{
Shipra Agrawal \thanks{Department of Computer Science, Stanford University. 
Email: shipra@stanford.edu. Research partially supported by Boeing.}
    \and 
Amin Saberi \thanks{Department of Management Science and Engineering, Stanford University. Email: saberi@stanford.edu}
\and
Yinyu Ye \thanks{Department of Management Science and Engineering,
Stanford University. 
Email: yinyu-ye@stanford.edu. Research partially supported by Boeing.}
} 
\maketitle
\date{ }
%-----------------------------------------------------
%   Abstract
%-----------------------------------------------------
\begin{abstract}
{\small
%Probabilistic constraints arise naturally in various applications involving uncertain problem data. Such constraints can be viewed as a compromise with the requirement of enforcing some constraints for all values of the uncertain data vector, which could be too costly or even impossible. 
In this paper, we present approximation algorithms for combinatorial optimization problems under probabilistic constraints. Specifically, we focus on stochastic variants of two important combinatorial optimization problems: the k-center problem and the set cover problem, with uncertainty characterized by a probability distribution over set of points or elements to be covered. 
%The goal is to minimize the covering cost (or covering distance) while ensuring that the probability with which the solution may fail to cover a random subset of elements is atmost a given input threshold. 
We consider these problems under adaptive and non-adaptive settings, and present efficient approximation algorithms for the case when underlying distribution is a product distribution. 
\newtext{In contrast to the {\it expected cost} model prevalent in stochastic optimization literature, our problem definitions support restrictions on the {\it probability distributions} of the total costs, via incorporating constraints that bound the probability with which the incurred costs may exceed a given threshold}. 
%with which a decision may fail or exceed a specified threshold.}
%\newtext{The approximation is in the cost of the solution while satisfying a constraint on the probability that the solution is feasible.}
%In contrast to some previous work on probabilistically constrained combinatorial optimization problems, our approximation algorithms do not incur any blowup of the probabilistic threshold. 
}
\end{abstract}
%-----------------------------------------------------
%  Introduction
%-----------------------------------------------------
\section{Introduction}
\label{Intro}
%A fundamental challenge in decision making is to deal with uncertain parameters and problem data while trying to achieve some predetermined objectives. Stochastic optimization provides a means to model the problem of determining optimal decision given one's assessment of the uncertain environment. Starting with the work of Dantzig \cite{dantzig} in the 1950s, these models have found increasing application in a wide variety of areas; see, e.g., \cite{birge, shapiro-survey}. Here, the uncertainty is modeled by a probability distribution over possible realizations of the actual data, called {\it scenarios}. Since the problem data is random, the objective can have many possible interpretations. A standard objective is to minimize expected cost over all the scenarios. However, the standard expected cost model is not always adequate to capture the stochastic setting. One such situation arises when the entire (optimal) decision must be taken prior to observing any random parameters. Often, one can hardly find any decision which would definitely satisfy the problem constraints 
%exclude constraint violation in all scenarios.
\newtext{A prevalent model to deal with uncertain data in optimization problems is to minimize expected cost over an input probability distribution. However, the expected cost model does not adequately capture the following two aspects of the problem. Firstly, in many applications, constraint violations cannot be modeled by costs or penalties in any reasonable way (e.g., safety relevant restrictions like levels of a water reservoir). Thus, if the problem constraints involve an uncertain parameter, one would rather insist on bounding the probability that a decision is infeasible.}
% That is, guaranteeing feasibility `as much as possible', that is, call a decision feasible if it is feasible `with high probability'. 
This leads to {\it probabilistic or chance constraints} of type:
$$P(h(x,\xi) \le 0) \ge 1-\rho$$  
	where $x$ and $\xi$ are decision and random vectors, respectively, ``$h(x,\xi) \le 0$" refers to a finite set of constraints, $P$ is a probability measure, and $\rho$ is a small input constant.

Another criticism of expected value measure is that it fails to capture the {\it risk} associated with the decisions: two decisions are valued equally if they have same expected cost. However, it can be the case that while one decision incurs moderate cost under all scenarios, the other incurs a huge cost for a disaster scenario with non-negligible probability. A risk averse user will naturally prefer the former decision. 
Various measures have been proposed in finance and stochastic optimization literature to capture this notion of risk averseness. A popular measure is the `value-at-risk (VaR)' measure, which is widely used in financial models, and has even been written into some industry regulations \cite{VaR1, VaR2}. For a given risk aversion level $\rho$, value-at-risk is given by the smallest value $\gamma$ such that probability that objective cost exceeds $\gamma$ is less than $\rho$. This leads to the probabilistic constraint:
$$P(f(x,\xi) \ge \gamma) \le \rho$$
where $f(x,\xi)$ is the objective value for decision $x$ in scenario $\xi$.

\newtext{In this paper, we develop approximation algorithms for such probabilistically constrained optimization problems.
% arising due to both the above situations. 
Specifically, we look at stochastic variants of two important combinatorial optimization problems: the k-center problem and the set cover problem, with uncertainty characterized by a probability distribution over subset of points or elements to be covered. We study the problems under ``non-adaptive" and ``adaptive" settings.
% Here, the locations or elements to be covered are uncertain; and all that is known is a probability distribution on the elements. The first situation arises in a 
In non-adaptive
%decision making 
setting, the entire set cover (k-center) must be chosen before the random element set is known. The goal is to minimize the covering cost (clustering distance) while satisfying a constraint that probability of covering a random subset of elements is higher than a given input threshold. In adaptive setting, the set cover (k-center) can be chosen adaptively for each scenario {\it after} observing the random element set. The goal is to determine the quality of optimal adaptive solution using value-at-risk (VaR) measure, that is, determine the minimum value $\gamma$ such that probability that the covering cost (clustering distance) exceeds $\gamma$ is less than $\rho$. Note that these two settings capture the two problem aspects mentioned in the previous paragraph.
%above.
}

Below we give formal definitions of our optimization problems and assumptions made on the statistical information available; followed by a summary of results and related previous work.
%-----------------------------------------------------
%  Problem Definitions
%-----------------------------------------------------
\paragraph{Non-adaptive stochastic k-center:} Consider a set $V$ of $n$ vertices. 
Assume that distance $d(u,v)$ between two vertices $u$ and $v$ in $V$ is given by a
graph metric $G=(V,E)$. The deterministic k-center problem is to find a subset $C \subseteq V$, $|C|\le k$, which minimizes the distance $r$ such that 
$$\max_{v\in V} d(v,C) \le r$$ 
In the stochastic k-center problem, the subset of $V$ that actually needs to be served is given by a random variable $\tilde{V}$, where each vertex $v_i$ appears in $\tilde{V}$ independently with probability $p_i$. 
The problem is to choose a set $C \subseteq V$, $|C|\le k$, which minimizes the distance $r$ such that 
$$P(\max_{v\in \tilde{V}} d(v,C) \le r) \ge 1-\rho$$ 
for a small input constant $0<\rho \le 1$.

\paragraph{Adaptive stochastic k-center}
In adaptive setting, the $k$ centers will be chosen after the random subset $\tilde{V}$ becomes known. Thus, the $k$-center solution $\tilde{C}$ is itself a random variable, and depends on the random subset $\tilde{V}$. The problem is to compute the value-at-risk, that is, the distance $r$ such that  
$$ P(\max_{v\in \tilde{V}} d(v,\tilde{C})> r) \le \rho$$
Here $\tilde{C}$ denotes optimal $k$-center solution for subset $\tilde{V}$.

\paragraph{Non-adaptive stochastic set cover}
Given a universe of $n$ elements $E=\{e_1, e_2, \ldots, e_n\}$, and a family ${\cal S}$ of $m$ subsets of $E$.
The deterministic set cover problem is to find the minimum cost subcollection $C \subseteq {\cal S}$ such that every element in $E$ is covered by some set in $C$. In the stochastic set cover problem, the elements to be covered are a random subset $\tilde{E}$ of $E$, where each element $e_j$ appears independently in $\tilde{E}$ with probability $p_j$. 
The problem is to find minimum cost sub-collection $C \subseteq {\cal S}$ such that the probability that every element in $\tilde{E}$ is covered is by some set in $C$ higher than an input threshold $1-\rho$.

\paragraph{Adaptive stochastic set cover}
In adaptive setting, the set cover will be chosen after the random subset of elements $\tilde{E}$ becomes known. 
The problem is to compute the value-at-risk $B$, that is the minimum value $B$ such that 
$$ P(\sum_{i \in \tilde{C}} c_i > B) \le \rho $$
Here $\tilde{C}$ denotes optimal set cover for random subset $\tilde{E}$.

%-----------------------------------------------------
%     Our results
%-----------------------------------------------------
\subsection{Summary of our results}
For the k-center problems (non-adaptive and adaptive), we present polynomial-time dynamic programming algorithms that give {\it optimal solutions} for {\it tree metrics}. Moreover, we show that the algorithms for tree metrics can be extended to give efficient PTAS for {\it planar graph metrics}, and more generally a class of graphs called `bounded genus' graphs. Here, the approximation is only in the number of centers; the probabilistic constraint holds exactly. For set cover problem, we give an $O(\log n)$-approximation algorithm for the non-adaptive case. We also show that for the adaptive case of this problem, verifying the probability threshold is atleast as hard as the problem of counting maximum independent sets of a graph, and hence is likely to be very hard to approximate. 

\newtext{We use combinatorial optimization techniques like dynamic programming to obtain fast and accurate algorithms for stochastic optimization problems. A common limitation of previous work \cite{goel99stochastic, swamy-risk-averse} on approximation algorithms for probabilistically constrained optimization problems is that the probabilistic constraint cannot be satisfied accurately. That is, an approximation of type $P(f(x,\xi) \ge (1+\epsilon_1) B) \le (1+\epsilon_2) \rho$ is obtained. We overcome this limitation by taking advantage of special structure of the problems in case of product distributions, and obtain approximation algorithms where probabilistic constraints hold exactly.}

%-----------------------------------------------------
%     Related Work
%-----------------------------------------------------
\subsection{Related Work}
There has been significant recent interest in studying stochastic optimization models from approximation algorithms perspective. A variety of approximation results have been obtained for the expected value models where a compensation or recourse is available for failed scenarios (the two-stage recourse models), see, e.g., \cite{swamy-survey} and references therein. 
But, results on probabilistically constrained and risk-averse models are relatively limited. In general, probabilistic constraints are difficult to handle: reasons being the inherent nonconvexity of the feasible set of a probabilistic constraint, as well as computational difficulty of estimating the probability term itself. However, some success has been achieved in obtaining approximation algorithms for specific combinatorial optimization problems by taking advantage of structure of the problem. In this regard, closest to our work are \cite{goel99stochastic} and \cite{cormode-pods07}.  
Goel et al \cite{goel99stochastic} focus on stochastic load balancing problems,
where item sizes are independent random variables following Bernoulli, exponential, or Poisson distributions specified in the input. They obtain approximation algorithms for stochastic bin packing and knapsack problems under probabilistic constraints that limit the overflow probability of a bin or knapsack.  These problems fall into the non-adaptive framework described above. Cormode et al \cite{cormode-pods07} emphasize the application of ``uncertain k-center" and similar clustering problems in probabilistic databases, and present various bi-criteria approximation (both in number of centers and maximum distance to a center) algorithms for an expected cost model. Their results include a $O(\log n)$ approximation in number of centers for our non-adaptive k-center problem, but only under an assumption that the individual probabilities $p_i$ are polynomial. As mentioned in the previous subsection, our work improves upon this result for tree and planar graph metrics. We give exact algorithm for tree metric case, and efficient PTAS for planar graphs, where the approximation factor and running time does not depend on the probabilities $p_i$. The adaptive k-center problem was not considered in the referred work.

A recent unpublished work by Swamy \cite{swamy-risk-averse} considers two stage risk-averse models for stochastic set cover and related combinatorial optimization problems. In the two stage recourse model, some sets can be chosen in the first stage at a low cost, and then if a scenario is not covered, more sets can be bought in the second stage as a recourse action. The risk averse problem is to minimize the sum of first stage cost and value-at-risk for the second stage. It was observed that if the value-at-risk for second stage is fixed to be $0$, the problem reduces to chance-constrained set cover without recourse - same as our non-adaptive set cover problem. Although the algorithms in \cite{swamy-risk-averse} can be used under more general assumptions of ``black box distributions", we present faster algorithms that achieve better approximation factors for the special case of product distributions. Specifically, {\it in contrast} to the results in \cite{swamy-risk-averse}, we do not incur any approximation in the probabilistic constraint, and the running time of our algorithms is independent of the input threshold $\rho$.  

An alternate approach widely used for dealing with probabilistic constraints focuses on replacing such constraints by more tractable constraints \cite{calafiore-chance06, neimro-chance06, iyengar-chance07} so that any solution satisfying the new constraints also satisfies the original probabilistic constraints with high probability. Observe that this type of relaxation is {\it opposite} to what one aims for in the design of approximation algorithms. 
Although some approximation results have been obtained \cite{calafiore-chance06, neimro-chance06, iyengar-chance07}, they apply only to problems involving continuous random variables whose distribution satisfies a certain concentration-of-measure property. These conditions are not fulfilled by problems with $0-1$ random vectors considered in this paper.

%-----------------------------------------------------
% stochastic k-center problem
% -- non-adaptive
% -- extensions to general graphs
%-----------------------------------------------------
\section{Non-adaptive stochastic k-center problem}
In this section, we look at the non-adaptive k-center problem. We present a dynamic programming algorithm for choosing a set $C\subseteq V$ of $k$ centers that maximizes the `success probability' $P(\max_{v\in \tilde{V}} d(v,C) \le r)$ for a given distance $r$. The final solution can then be found by doing binary search for optimal $r$ over a sorted list of ${n \choose 2}$ distances. 
%for maximizing the success probability $Pr(\max_{v\in \tilde{V}} d(v,C) \le r)$  for a given $r$. 
Below, we first describe an exact algorithm for tree metrics. The algorithm is similar in spirit to the dynamic programming algorithm given in \cite{p-median} for (deterministic) $k$-median problem under tree metrics. 
%The algorithm is exact when distances are given by a tree metric $T=(V,E)$.
In the sequel, we extend this algorithm to obtain approximation algorithms for more general graph metrics.
%The optimal solution is given by the minimum value of $r$ for which the success probability meets the threshold of $1-\rho$.
%Note that this procedure requires an $O(n^3)$ preprocessing step that computes all pair shortest paths.

\subsection{Exact algorithm for tree metrics}
Our algorithm for tree metrics is based on a key property of our model, that is, ``for any subtree in a tree, once the number of centers in the subtree and the center closest to its root are fixed, the probability of success for the subtree is independent of the rest of the tree". The reason this property holds lie with the structural properties of the problem on a tree graph, and our independence assumption on probability of vertices. The hierarchical structure of tree ensures that the closest center to any vertex in a subtree either lies inside the subtree or is the center closest to the root of the subtree. 
The independence assumption on vertices implies that inter-dependencies between disjoint subtrees are caused only due to the common centers used to cover them. Once the closest center to root and number of centers in the subtrees are fixed, the joint probability of success for a tree can be expressed as product of success probabilities for its subtrees.  This observation will give us the optimal substructure property required for a dynamic programming approach.

We make these ideas more precise in the following.
 
\paragraph{Dynamic programming algorithm}
Given a rooted tree $T=(V, E)$ with root $v_0$. $T_{v}$ denotes the subtree of $T$ under vertex $v$ (including $v$), $e(v,t)$ denotes the $t^{th}$ child edge of vertex $v$, and $T_{e(v,t)}$ denotes the subtree of $T_v$ on the left of the edge $e(v,t)$ (including $v$ and edge $e(v,t)$). 
Also, $t_s$ denotes the total number of child edges of a vertex $v_s$.

%We first define some notations
%\begin{itemize}
%\item $v_0$ denotes the root of tree $T$.
%\item $T_{v}$ denotes the subtree of $T$ under a vertex $v$ (including $v$).
%\item $e(v,t)$ denotes the $t^{th}$ child edge adjacent to $v$ in $T_v$.
%\item $t_s$ denotes the total number of children of $v_s$ in $T_{v_s}$
%\item $T_{e(v,t)}$ denotes the subtree of $T_v$ on the left of the edge $e(v,t)$ (including $v$ and edge $e(v,t)$). $T_{e(v,0)}$ denotes the vertex $v$.
%\item $Lev(v)$ denotes the level of vertex $v$ in tree $T$. Level of the root of the tree is $0$.
%\item $L$ is the maximum depth of the tree $T$.
%\item For any pair of vertices $v_1$ and $v_2$, $p(v_1,v_2)$ denotes the unique path between them.
%\end{itemize}

Now, for any subtree $\bar{T}=\{S,\bar{E}\}$ of $T$, define function $H(\bar{T}, j)$ as maximum probability (i.e., the probability under optimal choice of centers) that random subsets $\tilde{S}$ of $S$ can be {\it covered} by $j$-centers. Given clustering distance $r$, we say that a set of vertices is {\it covered} by a set of centers iff for every vertex there is some center within distance $r$.
$$ H(\bar{T},j) = \max_{C_j \subseteq S, |C_j|=j } Pr(C_j \mbox{ covers } \tilde{S})$$
Note that $H(T,k)$ gives the desired optimal value. % objective value. 
We now define function $R(\bar{T}, j, v)$ which will prove to be an essential tool for computing values of $H(\cdot)$. Suppose it is given that $v$ is a closest center to the root of the subtree $\bar{T}$, then $R(\bar{T}, j, v)$ is defined as maximum probability that a set of $j-1$ centers in $S$, along with the center $v$, can cover a random subset of $S$. That is,
$$R(\bar{T}, j, v) = \max_{%\begin{array}{c}
			C_{j-1}\subseteq S, |C_{j-1}|=j-1
			%\end{array}
		}Pr(C_{j-1} \cup v \mbox{ covers } \tilde{S})$$

We employ a dynamic programming type procedure that proceeds bottom up in the tree and computes all values of $R(T_{e(v_1,l)}, j, v_2)$ and $H(T_v, j)$ (and finally $H(T,k)$ for the whole tree $T$). 
\paragraph{The initial values:} For any leaf $v$, $T_v =v$,
$$H(T_{v},j) = \Bigg\{\begin{array}{cc} 
			1 & \mbox{ if } j\ge 1 \\
			1-p_v & o.w.	
			\end{array}
$$
Also, for any vertex $v_1$, $T_{e(v_1,0)}=v_1$. So, for any pair of vertices $v_1,v_2$:
$$R(T_{e(v_1,0)},j,v_2) = \displaystyle\Bigg\{\begin{array}{ll} 
			1 & \mbox{ if } j > 1\\
			1 & \mbox{ if } j = 1, d(v_1,v_2) \le r\\
			1-p_1 & o.w. 
			\end{array}
$$

\begin{figure}[htbp]
\begin{center}
\includegraphics[totalheight=0.17\textheight]{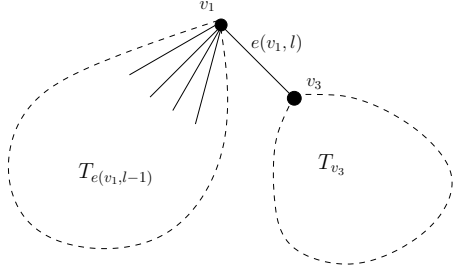}
\caption{Tree $T_{e(v_1,l)}$ and its subtrees}
\label{fig1}
\end{center}
\end{figure}

%\begin{wrapfigure}{r}{0.4\textwidth}
%\centering
%\includegraphics[width=0.4\textwidth]{dp1.ps}
%\caption{Text wrap around figure}
%\noindent \hrulefill
%\label{test}
%\end{wrapfigure}
\noindent {\it \bf Computation of $H(T_{v_1},j)$: }
Let $C$ be the optimal set of $j$-centers for tree $T_{v_1}$, and $v_r$ be the closest center to $v_1$ in $C$. Then, by the definition: 
$$H(T_{v_1},j) =R(T_{e(v_1, t_1)}, j, v_r)$$
Therefore we can compute $H(T_{v_1},j)$ using the following relation:
$$H(T_{v_1},j) = \max_{v_2 \in T_{v_1}} {R(T_{e(v_1,t_1)}, j, v_2)}$$

%\noindent {\it \bf 
\paragraph{Computation of $R(T_{e(v_1,l)},j, v_2)$:}
By definition, $v_2$ is closest vertex to the root $v_1$ of the subtree; and $l\le t_1$, the number of child edges of $v_1$. If $v_1$ is a leaf, then $t_1=0$, and $R(e(v_1,0),j,v_2)$ is given by the initial values. 
Assume that $v_1$ is not a leaf and $l \ge 1$. Let $v_3$ be the vertex on the other end of edge $e(v_1,l)$ (refer Figure \ref{fig1}). The value of $R(T_{e(v_1,l)},j, v_2)$ is given by the following recursion:
$$
\begin{array}{l}
R(T_{e(v_1,l)},j,v_2)= \max_{j_1, j_2 \in [0,j]} \{ R(T_{e(v_1,l-1)},j_1,v_2) \cdot R(T_{e(v_3,t_3)},j-j_1+1,v_2),\\
\hspace{3in} R(T_{e(v_1,l-1)},j_2,v_2) \cdot H(T_{v_3},j-j_2)\}
\end{array}
$$

The reason this equation holds is as follows. 
Since $v_2$ was the closest center to the root of subtree $T_{e(v_1,l)}$, it remains closest center to the root of subtree $T_{e(v_1,l-1)}$. However for subtree $T_{v_3}$ (same as $T_{e(v_3,t_3)}$), there are two possible choices: either $v_2$ remains the closest center, or a center in $T_{v_3}$ is the closest center.  
The two terms on the right represent these two choices. The product expression follows from the independence property discussed in the beginning of this section.
 
We order the vertices of the tree from bottom to top and left to right. 
At stage $i$, we compute values $R(T_{e(v_1,l)},j,v_2)$ for $i^{th}$ vertex $v_1$ picked in this order. For a given vertex $v_1$, $R(T_{e(v_1,l)},j,v_2)$ is computed for increasing values of $l$ and $j$, and all choices of $v_2$ in $T$. Then, we compute values $H(T_{v_1},j)$, and go on to the next stage.  
Thus, at any stage, all the terms in above expression are already known from computations in the previous stages. 

\paragraph{Computing the optimal solution}
Assume that we have calculated (and recorded) all values of $H(\cdot)$ and $R(\cdot)$. $H(T,k)$ gives the optimal probability. The corresponding optimal set of $k$-centers can be generated by carrying out another pass over this table of values. This is a standard component of any dynamic programming procedure, we omit the details here. 

\paragraph{Running time complexity} For each edge $e(v_1,l)$ and each vertex $v_2$, we compute $R(T_{e(v_1,l)},j,v_2)$ for all $k$ values of $j$. 
Also, each computation of $R(\cdot)$ requires taking $\max$ over atmost $2k$ terms. Therefore total complexity of computing
the terms $R(\cdot)$ is $O(n^2k^2)$.
For each vertex $v$, there are atmost $k$ values of $j$ for which $H(\cdot)$ need to be computed. 
And each of these computations takes $O(n)$ steps. Hence, total complexity of computing terms $H(\cdot)$ is $O(n^2k)$. 

Also, as a pre-procedure for the algorithm we compute the distance-matrix of the tree (this requires $O(n^2)$ steps). And, the algorithm needs to be repeated for $\log{n^2}$ possible values of $r$. Thus, total complexity of the procedure is $O(n^2k^2\log n)$.

\subsection{Extensions} 
\paragraph{Extensions to more general graph metrics}
In this section, we extend our algorithm to obtain efficient PTAS
for planar graphs and a more general class of graphs called ``bounded genus graphs". The heart of this approach lies in the adaptability of the structure of {\it c-outerplanar graphs} to dynamic programming. A c-outerplanar graph has the property that it can easily be decomposed into two subgraphs with just 2c common boundary nodes \cite{baker94}. 
Now, a dynamic programming algorithm similar to our algorithm for tree case can be used. For a given $c$, let $G$ is a $c$-outerplanar graph. Then, using techniques in \cite{baker94}, $G$ can be recursively decomposed into $c$-outerplanar subgraphs $G_1$ and $G_2$ with atmost $2c$ common boundary nodes. The dynamic programming recursion is now defined as:
\[
	%H(G, j) = \max_{v_1, v_2 \in C} R(C, j, v_1, v_2)
	H(G, j) = \max_{\{v_i\}_{i=1}^{2c} \subseteq V} R(G, j, \{v_i\})
\]
$$\begin{array}{l}
	%R(G, j, v_1, v_2) = \max_{0 \le j_1, j_2 \le j} \{ R(G_1, j_1, v_1, v_2) \cdot R(G_2, j-j_1+1,v), R(G_1, j_1, v_1, v_2) \cdot R(G_2, j-j_1+1,v_2), \\
%\hspace{0.2in} R(G_1, j_1, v_1, v_2) \cdot R(C_2, j-j_1+2,v_1, v_2), R(G_1, j_2, v_1, v_2) \cdot H(C_2, j-j_2) \}
	R(G, j, \{v_i\}) = \max_{0 \le j_1, j_2 \le j} \Big\{ \max_{U\subseteq \{v_i\}, U \ne \phi}  R(G_1, j_1, \{v_i\}) \cdot R(G_2, j-j_1+|U|,U), \\
					\hspace{2in}R(G_1, j_2, \{v_i\}) \cdot H(G_2, j-j_2) \Big\}
\end{array}
$$
%where $G_1$ and $G_2$ are two subgraphs of $G$ with $2c$ common boundary nodes.
Since there are $n$ vertices, there are atmost $nk$ values of $G$ and $j$ for which $H$ has to be computed, and each computation requires taking max over $n^{2c}$ values. So complexity of computing terms $H(\cdot)$ is $O(n^{2c+1}k)$. Similarly, there are $n^{2c+1}k$ values for which $R(\cdot)$ has to be computed. Each computation requires taking max over $2^{2c+1}k$ terms. %So complexity of computing terms $H(\cdot)$ is $O(n^3k^2)$. 
Hence, total running time complexity of above procedure is $O(n^{2c+1}k^2)$. 

To extend this approach to general planar graphs, we can use graph decomposition concepts from \cite{baker94}. Here, we give an outline of the method. 
%We will get the optimal distance $r$ at the cost of a small constant approximation in $k$. 
The idea is to decompose the planar graph into disjoint $(c+1)$-outerplanar components by copying the nodes in every $c^{th}$ `level' \cite{baker94}. Then, use the above algorithm for resulting $(c+1)$-outerplanar graph. Note that we are potentially duplicating the centers in the copied levels. However, by pigeonhole principle, there exists $i\in \{1,\ldots, c\}$ such that if we copy levels congruent to $jc+i$, $j>0$, then number of centers increase by a factor of atmost $1+1/c$. This gives a $(1+1/c)$-approximation in number of centers, with running time $\tilde{O}(n^c)$.
A result by Eppstein \cite{eppstein} shows that similar decompositions can be achieved in polynomial-time for a more general class of graphs called ``bounded genus" graphs. Thus, our approximation algorithms extend in a natural way to this class of graphs. 

\paragraph{Extensions to other covering problems}
Our algorithm can be directly applied to other stochastic covering problems on planar graphs, like vertex cover, edge cover and dominating set. The basic idea remains the same: once we fix the number of centers (covering nodes or edges) in a subgraph and the closet center(s) to its boundary node(s), the probability of covering the subgraph is independent of the rest of the graph. Note however, that for problems with non-uniform cost of centers, our dynamic programming algorithm will be pseudo-polynomial (polynomial in `total cost').
%However, for problems with non-uniform cost of centers, some more consideration is needed. Specifically, in the expression for computing values of function $R(\cdot)$, we take maximum over all possible divisions of the total cost of centers. In case of non-uniform costs, in order for this operation to be polynomial-time, we need to assume that the total cost is polynomial in $n$.
%-----------------------------------------------------
% stochastic k-center problem
% -- adaptive
%-----------------------------------------------------
\section{Adaptive stochastic k-center problem}
In adaptive setting, the goal is to find the minimum distance $r$ such that the {\it failure probability} $P(\max_{v \in \tilde{V}} d(v,\tilde{C}) > r)$ is less than $\rho$. Again, the desired value $r$ could be found by doing a binary search over ${n \choose 2}$ values of $r$, and testing for each $r$ whether the failure probability is less than $\rho$. 
%so that with high probability (greater than $1-\rho$), a random subset of vertices has a $k$-center solution with maximum coverage distance $r$.  Again, the desired value $r$ could be found by iterating over $n^2$ values of $r$, and testing for each $r$ whether the {\it failure probability} $Pr(\max_{v \in \tilde{V}} d(v,\tilde{C}) > r)$ is less than $\rho$.  
However, evaluating this probability term is not straightforward. Here, a key difference from the non-adaptive setting is that a different set of centers $\tilde{C}$ is chosen for each random scenario $\tilde{V}$, optimized for the subset of vertices in that scenario. A brute force approach to find the failed scenarios would require solving a deterministic $k$-center problem for each of the $2^n$ subsets of $V$.
%A brute force approach would require solving a deterministic $k$-center problem for each of the $2^n$ subsets of $V$, in order to find the failed scenarios.
%out if that scenario fails to satisfy the covering condition $\max_{v \in \tilde{V}} d(v,\tilde{C}) \le r$; and then adding up the probabilities of the scenarios that do fail.

%\noindent Observe that the above problem can be solved by iterating over $k$ starting with $k=0$, and computing the failure probability $Pr(\tilde{V} \textrm{ is not covered by } f(k,\tilde{V}, E))$. If this probability is less than $p$, the corresponding $k$ will give the optimal budget $ck$. However, computing the failure probability term is not straightforward, since it involves evaluating exponential number of scenarios (subsets $\tilde{V}$ of $V$), and a k-center problem needs to be solved for each scenario (to find if $\tilde{V}$ is covered by $f(k,\tilde{V}, E)$). In the sequel, we present a polynomial-time dynamic programming algorithm for the tree case. As before, it can be extended to the planar graphs and gives an approximation scheme for general graphs.
%\noindent 
%\paragraph{Algorithm} 
In this section, we propose a dynamic programming algorithm to compute this failure probability in polynomial-time for a given value of $r$. 
%The algorithm gives an optimal solution in polynomial time for the case when distance between vertices is given by a tree metric $T=(V, E)$. 
First, we present an exact algorithm for tree metrics, and then extend it to more general graph metrics.

\subsection{Exact algorithm for tree metrics}
The basic idea in our algorithm is to characterize each random subset of a subtree via a profile $(j,d,d')$ that completely captures its covering properties. Specifically, given a subtree $\bar{T}=\{S,\bar{E}\}$, a random subset $\tilde{S}\subseteq S$ belongs to a profile $(j,d,d')$ if and only if
\begin{itemize}
\item the {\it minimum} number of centers sufficient to cover $\tilde{S}$ within distance $r$ is $j$,
\item among the covers of size $j$, minimum distance of a center to the root of $\bar{T}$ is $d$, and
\item $d'$ is the maximum distance such that if a vertex $v'$ outside the subtree $\bar{T}$ and at distance $d'$ from its root is a center, then the subtree can be covered using only $j-1$ centers. 
%If there is no such vertex $v_r$ outside the subtree, then $d'=-d$. 
%\item $d'$ is the maximum distance of a vertex $v_r$ {\it outside} the subtree $\bar{T}$ so that if an additional center is provided at a vertex $v_r$, then $j-1$ centers in the subtree are sufficient to cover $\tilde{S}$. 
If no such vertex $v'$ exists, then $d'=-d$. 
\end{itemize}
Note that each subset of vertices belongs to exactly one profile $(j, d, d')$. This is because there is a unique minimum
number of centers $j$ required for any subset, and that corresponds to a unique minimum distance $d$ of closest center
to the root. Also note that using any {\it help} $v'$ from 
outside the tree atmost $1$ center can be removed out of the $j$ centers -- otherwise we could place a center at the root and reduce the minimum  number of centers to $j-1$. Taking maximum of the distances of all such $v'$s from root, we get our unique $d'$. 

Above argument shows that the profile $(j,d,d')$ define a disjoint partition over the subsets of any subtree $\bar{T}$. 
Now, define function $DP(\bar{T},j,d,d')$ as the probability of random subsets in $\bar{T}$ under profile $(j,d,d')$. Then, by definition, the probability of failure is given by:
\begin{eqnarray}
\mbox{Failure probability} & = & \sum_{k<j\le n, d} DP(T, j, d, -d) 
\end{eqnarray}
Here, $d$ can take atmost $n$ possible values -- corresponding to possible distances of vertices from the root.
%of vertices of the tree from its root. 
Now, we are ready to present our dynamic programming algorithm. We use the same notations as in the previous section. The algorithm will compute all values $DP(T_{e(v_1,l)},j, d, d')$ in a bottom to top, left to right order, finally computing the values $DP(T, j, d, -d)$ that appear in the above expression for failure probability.

\paragraph{Initial values: } For $l=0$, $T_{e(v,0)}=v$,
$$
\begin{array}{l} 
DP(v, j, d, d')  = \Bigg\{ \begin{array}{cl}
				p_v & \mbox{if } j=1, d=0, d'=\max_{v'\ne v, d(v,v')\le r} d(v,v') \\
				1-p_v & \mbox{if } j=0, d=0, d'=0\\
				0  & o.w.  	
				\end{array}
\end{array}
$$

\paragraph{Computation of $DP(T_{e(v_1,l)}, j, d, d')$: }
Now, assume that $v_1$ is not a leaf, and $l\ge 1$. 
To compute $DP(T_{e(v_1,l)}, j, d, d')$ for some $l$, we reduce it to an expression consisting of function $DP(\cdot)$ on subtrees $T_1=T_{e(v_1,l-1)}$ and $T_2=T_{e(v_2,t_2)}$, where $v_2$ is the vertex on the other end of edge $e(v_1,l)$. We use the observation that a random subset $\tilde{V}$ 
of this tree has a profile $\{j,d,d'\}$ if and only if $\tilde{V_1} = T_1 \cap \tilde{V}$ 
and  $\tilde{V_2} = T_2 \cap \tilde{V}$ have profiles $(j_1,d_1,d_1')$ and $(j_2,d_2,d_2')$, respectively, satisfying {\it either} of the following conditions:
% are satisfied:
%that $0 \le j_1, j_2 \le n$, $d_1 \in D(T_1), d_2 \in D(T_2)$, $d_1' \in D'(T_1), d_2' \in D'(T_2)$ and either of the following conditions is satisfied:
\begin{itemize}
\item $j_1+j_2=j$: In this case, we must ensure that the centers in $V_1$ do not help $V_2$ and vice-versa so that total 
minimum number of centers is $j$. Let $w$ denote the distance $d(v_1,v_2)$, then we require $d_2+w>d_1', d_1+w>d_2'$. To get $d$, the least of $d_1$ and $d_2+w$ must be equal to 
$d$, and to get $d'$, the max of $d_1'$ and $d_2'-w$ must be equal to $d'$.
% d_2+w>d_1' d_1+w>d_2', d_1'=d', d_2'<d'+w, d_1=d, d_2+w>d
\item $j_1+j_2=j+1$: In this case we must ensure that the centers in $V_1$ help $V_2$ {\it or} vice-versa, so that total 
minimum number of centers is $j$, that is $d_2 + w \le d_1', d_1 + w>d_2'$ or $d_2+w>d_1', d_1+w\le d_2'$. To get $d$, the least of 
$d_1$ and $d_2+w$ must be equal to $d$. To get $d'$, $d_1'$ must be equal to $d'$ if $V_2$ is helped by $V_1$, and $d_2'-w$ 
must be equal to $d'$ if $V_1$ is helped by $V_2$.
\item $j_1+j_2=j+2$: In this case we must ensure that the centers in $V_1$ help $V_2$ {\it and} vice-versa, so that total
minimum number of centers is $j$, that is $d_2\le d_1'-w, d_1 \le d_2'-w$. To get $d$, the least of 
$d_1$ and $d_2+w$ must be equal to $d$. Only negative values of $d' (=-d)$ have this case.
\end{itemize}
It is easy to see that in each of the above cases, the conditions on $d_1,d_2$ and $d'_1, d_2'$ are necessary and sufficient to get the joint profile $(j,d_1,d_2)$. Let $\cal P$ denotes the collection of profiles $\{(j_1,d_1,d_1')$, $(j_2,d_2,d_2')\}$ satisfying either of the above conditions. 
Then, using the fact that the profiles are disjoint, and independence assumptions on the probability model, $DP(T_{e(v_1,t)}, j, d, d')$  can be expressed as
$$
\begin{array}{l} 
DP(T_{e(v_1,l)}, j, d, d') = \sum_{\cal P} DP(T_{e(v_1,l-1)}, j_1, d_1, d_1') \cdot DP(T_{e(v_2,t_2)}, j_2, d_2, d_2')
\end{array}
$$
%DP(e(v_1,j), k, d, d-) =     first set of conditions ensure that minimum number of centers is k. Second set ensures that d- is the minimum distance of extra center required to get k-1. Third set ensures that d is the minimum distance of closest center to root
%
%1	         sum{k1+k2=k, d2+e>d_1' d_1+e>d2', d1'=d', d2'<d'+e, d1=d, d2+e>d}
%2		 .............................., d1'<d', d2'=d'+e, ..................
%3		 .............................., d1'=d', d2'<d'+e, d1>d, d2+e=d 			
%4		 .............................., d1'<d', d2'=d'+e, ..................
%		  
%2,4 d2'=d'+e 	       sum{k1+k2=k+1, d2+e<=d1' d1+e>d2' 
%1,3 d1'=d' 	       sum{k1+k2=k+1, d2+e>d1' d1+e<=d2' 
%		all combinations
%		DP(e(v_1,j-1), k1, d1, d1-) DP(e(v_2,t2), k2, d2, d2-)
%Assume that we compute $DP(\cdot)$ for vertices $v_1$ from leaves to higher up in the tree. And, for a given $v_1$, we compute $DP(\cdot)$ for edges $<t$ before computing it for $t$th edge. 
Observe that due to the specific order in which we compute the values of $DP(\cdot)$, all terms in the above expression were already computed in a previous stage. 

\paragraph{Running time complexity}
For each edge, we compute atmost $kn^2$ values of $DP(\cdot)$ (possible values of $j$ and $d$, $d'$). For each of these 
terms we sum over at most $3kn^4$ terms. Therefore, total complexity is $O(k^2n^6)$. The preprocessing time is $O(n^3)$ for computing distance pairs, and $O(n^2)$ for assigning initial values. Including the $\log n^2$ iterations for binary search on $r$, the effective complexity is $O(k^2n^6\log{n})$. 
%the algorithm needs to be repeated for each of the $n^2$ possible values of $r$. Therefore, the effective complexity is $O(k^2n^9)$.

\subsection{Extensions}
The algorithm can be extended to more general graph classes and other covering problems on graphs, using ideas similar to those discussed at the end of previous section. We omit the details here.
%\shcomment{TODO}
 %-----------------------------------------------------
% stochastic set-cover problem
% -- non-adaptive
%-----------------------------------------------------
\section{Non-adaptive stochastic set cover problem}
We give an approximation method for non-adaptive stochastic set cover problem by reformulating it as a partial set cover problem. The problem (refer Section \ref{Intro}) can be restated as:
\begin{eqnarray*}
\label{set-cover-1-stage}
\min_{x} & \sum_{i=1}^m c_ix_i & \nonumber\\
\textrm{s.t. } & P(\tilde{E} \textrm{ is not covered by } x) \le \rho & \nonumber\\
	& x_i \in \{0,1\} & \forall i \in [n]
\end{eqnarray*}
 Here, $[n]$ denotes the set $\{1, \ldots, n\}$. The value of 0-1 variable $x_i$ indicates whether set $i$ is chosen or not. 

For any element $j$, let $\partial{j}$ denote the collection of sets that cover the element $j$. Then, indicator function $I_j(x) = (1-\sum_{i\in \partial{j}} x_i)^+$ takes value $1$ if $j$ is NOT covered by solution $x$ and $0$ otherwise. 
Using the assumptions on our probability model:
%The probability of failure for a solution $x$ is then given by
%Then, any element in $D$ is covered if and only if atleast one of the sets that cover it appears in $S$. 
\begin{eqnarray*}
P(\tilde{E} \textrm{ is not covered by } x) & = & 1-\Pi_{j:I_j(x)} (1-p_j)
\end{eqnarray*}

Let $l_j = \log{\frac{1}{1-p_j}}$, and $l=\log{\frac{1}{1-\rho}}$. Then, the probabilistic constraint is equivalent to:
%$P(\tilde{E} \textrm{ is not covered by }x) \le \rho$ is equivalent to:
\begin{eqnarray*}
\label{log-constraint}
1-\Pi_{j:I_j(x)} (1-p_j) & \le & \rho \nonumber\\
\Leftrightarrow \Pi_{j:I_j(x)} \frac{1}{1-p_j} & \le & \frac{1}{1-\rho}\nonumber\\
\Leftrightarrow \sum_{j=1}^n I_j(x) l_j & \le & l\nonumber
\end{eqnarray*}
Therefore, we can reformulate our problem as:
\begin{eqnarray*}
\min_{x} & \sum_{i=1}^m c_ix_i & \nonumber\\
\textrm{s.t. } & \sum_{j=1}^n  (1-\sum_{i\in \partial{j}} x_i)^+ \cdot l_j \le l & \nonumber\\
	& x_i \in \{0,1\}
\end{eqnarray*}
which is equivalent to the following integer program:
\begin{eqnarray}
\min_{x} & \sum_{i=1}^m c_ix_i & \nonumber\\
\textrm{s.t. } & \sum_{i\in \partial{j}} x_i \ge 1-z_j & \forall j=1,\ldots,n \nonumber\\
& \sum_{j=1}^n  z_jl_j \le l & \nonumber\\
	& x_i \in \{0,1\} & \forall i=1,\ldots,m \nonumber\\
	& z_i \in \{0,1\} & \forall j=1,\ldots,n \nonumber
\end{eqnarray}

The above problem can be interpreted as a `partial set cover problem', where penalty for not covering an element $j$ is given by $l_j$. The partial set cover problem is to minimize the cost of sets ($c^Tx$) such that the total penalty ($\sum z_jl_j$) for uncovered elements is less than a given limit ($l$). A $(\frac{4}{3}+\epsilon) \log n$-approximation algorithm for the partial set cover problem is appears in \cite{partialCover}. The algorithm can be directly used for the above problem. 

%-----------------------------------------------------
% stochastic set-cover problem
% -- adaptive
%-----------------------------------------------------
\section{Adaptive stochastic set-cover problem}
In the adaptive setting, our goal is to compute the minimum value $B$ so that probability that cost of optimal set cover for a random subset of elements in $E$ exceeds $B$, is less than $\rho$. Given a fixed value $B$, we call the subsets of $E$ with adaptive set cover cost $>B$ as {\it failed subsets}, and probability of these subsets as {\it failure probability}. We show that even for the uniform cost edge cover case, the problem of {\it approximating} this failure probability is harder than the problem of approximately counting maximum independent sets in a graph. 
An inapproximability result for the latter problem appears in \cite{luby-counting}, which states that this problem cannot be approximated within a polynomial factor unless RP=NP (refer Theorem $4$ in \cite{luby-counting}). 
%The literature on approximate counting suggests that the problem of counting independent sets is hard to approximate in general. 
%\shcomment{look for results on counting maximum independent sets}
%Dyer et al \cite{dyer} show that any Markov Chain Monte Carlo technique is likely to fail, and no fully randomized approximation scheme exists for dense graphs. 
Thus, a reduction from this problem will suggest that our problem is hard to approximate as well.

Given graph $G=(V,E)$ and a parameter $k$, we denote the edge cover failure probability by $f(k)$. It is the probability of random subsets $\tilde{V}$ of $V$ such that the number of edges in the edge cover of $\tilde{V}$ is greater than $k$. We call such subsets of $V$, the ``failed subsets". Let each vertex 
appears independently in the random subset $\tilde{V}$ with probability $p$ (that is, $p_i=p$ for all $i$). 
Denote by $N_i(G,k)$ the number of failed subsets containing $i$ vertices. Then,
$$f(k)=\sum_{i=k}^n N_i(G,k) \cdot p^{i} (1-p)^{n-i} $$
Denote the count of maximum independent sets of graph $G$ by $I(G) (\ge 1)$. 
 Let $m$ be the size of a maximum independent set in $G$.
We show that computing $f(m-1)$ with a good approximation factor is harder 
than approximating the number of independent sets $I(G)$. Note that $N_m(G,m)$ denotes the number of subsets of $V$ that have $m$ vertices and need $m$ or 
more edges to cover them. The edge cover needs $m$ or more edges to cover $m$
vertices if and only if the $m$ vertices form an independent set. Hence, 
$N_m(G,m) = I(G)$. Therefore,
\begin{eqnarray}
\label{lower-bound}
f(m) & = & I(G)p^m (1-p)^{n-m} + \sum_{i=m+1}^n N_i(G,m) p^{i} (1-p)^{n-i} \nonumber\\ 
& \ge & I(G)p^m (1-p)^{n-m}
%&&\rho(m) \nonumber\\
%&&= I(G)p^m (1-p)^{n-m} + \sum_{i=m+1}^n N_i(G,m) p^{i} (1-p)^{n-i} \nonumber\\ 
%&&\ge I(G)p^m (1-p)^{n-m}
\end{eqnarray}
Observe that $\sum_{i=m+1}^n N_i(G,m) \le 2^n \le 2^n I(G)$. Also, assume 
$p\le1/2^n$. Then,
\begin{eqnarray}
\label{upper-bound}
f(m) & = & I(G)p^m (1-p)^{n-m} + \sum_{i=m+1}^n N_i(G,m) p^{i} (1-p)^{n-i} \nonumber\\
%&& \rho(m) \nonumber\\
%&& \le I(G) p^m (1-p)^{n-m} + \sum_{i=m+1}^n N_i(G,m) p^{m+1} (1-p)^{n-m-1} \nonumber\\
& \le & I(G) p^m (1-p)^{n-m} (1+ 2^n I(G)\frac{p}{1-p}) \nonumber\\
& \le & I(G) p^m (1-p)^{n-m} ( 1+ 2^n \cdot \frac{1}{2^n-1}) \nonumber\\
& \le & I(G) p^m (1-p)^{n-m} (1+2)
\end{eqnarray}
From inequalities (\ref{lower-bound}) and (\ref{upper-bound}), we can conclude that 
\[\frac{1}{3} \cdot \frac{f(m)}{p^m (1-p)^{n-m}} \le I(G) \le \frac{f(m)}{p^m (1-p)^{n-m}}\]
Thus, if we have a $(1\pm \epsilon)$ approximation of $f(m)$, then we could
get a $(1\pm (\epsilon+\frac{2}{3}))$ approximation for $I(G)$. These completes the reduction.
%This proves that it is hard to get an algorithm that approximates $\rho(k)$. 
%We have assumed here that $p\le \frac{1}{2^n}$. The problem is still 
%\#P-hard and cannot
%be approximated by brute force method if $\rho(k) \le \frac{1}{2^{n^2}}$. 

%\let\oldthebibliography=\thebibliography
%  \let\endoldthebibliography=\endthebibliography
%  \renewenvironment{thebibliography}[1]{%
%    \begin{oldthebibliography}{#1}%
%      \setlength{\parskip}{0.3ex}%
%      \setlength{\itemsep}{0.99ex}%
%  }%
%  {%
%    \end{oldthebibliography}%
%  }

\bibliographystyle{plain}
\bibliography{paper}

\end{document}